\documentclass[aps,prl,reprint,showpacs,showkeys,superscriptaddress,preprintnumbers,nofootinbib]{revtex4-1} 
\pdfoutput=1

\usepackage{amsmath,amssymb,bm,mathrsfs}
\usepackage{hyperref}
\usepackage{url}
\usepackage{breakurl}
\usepackage{graphicx}

\newcommand{\SU}{ \mathrm{SU} }

\def\simgt{\mathrel{\lower2.5pt\vbox{\lineskip=0pt\baselineskip=0pt
           \hbox{$>$}\hbox{$\sim$}}}}
\def\simlt{\mathrel{\lower2.5pt\vbox{\lineskip=0pt\baselineskip=0pt
           \hbox{$<$}\hbox{$\sim$}}}}

\begin{document}

\title{Electroweakly-Interacting Dirac Dark Matter}

\author{Natsumi Nagata}
\affiliation{William I. Fine Theoretical Physics Institute, School of
Physics and Astronomy, University of Minnesota, Minneapolis, MN 55455, USA}
\affiliation{Kavli Institute for the Physics and Mathematics of the
  Universe (WPI) Todai Institutes for Advanced Study, University of
  Tokyo, Kashiwa 277-8583, Japan}
\email{natsumi.nagata@ipmu.jp}

\author{Satoshi Shirai}
\affiliation{Deutsches Elektronen-Synchrotron (DESY), 22607 Hamburg, Germany}
\email{satoshi.shirai@desy.de}

\begin{abstract}
 We consider a class of fermionic dark matter candidates that are
 charged under both the SU(2)$_L$ and U(1)$_Y$ gauge interactions.
 In this case a certain amount of dark matter-Higgs couplings, which can
 split the dark matter into a pair of Majorana fermions, should be
 present to evade the constraints from the dark matter direct detection
 experiments. These effects may be probed by means of the dark
 matter-nucleus scattering via the Higgs-boson exchange process, as well
 as the electric dipole moments induced by the dark matter and its
 SU(2)$_L$ partner fields. In this article, we evaluate them with
 an effective field approach. It turns out that the constraints coming from the 
 experiments for the quantities have already restricted the
  dark matter with hypercharge $Y\geq 3/2$. Future experiments
 have sensitivities to probe this class of dark matter candidates, and
 may disfavor the $Y\geq 1$ cases if no signal is observed. In this case,
 only the $Y=0$ and $1/2$ cases may be the remaining possibilities for
 the SU(2)$_L$ charged fermionic dark matter candidates. 
\end{abstract}

\maketitle
\preprint{DESY 14-201}
\preprint{FTPI-MINN-14/38}
\preprint{IPMU14-0332}

\section{Introduction}

Weakly interacting massive particles (WIMPs) are well-known candidates for dark
matter (DM) in our Universe. The thermal relic abundance of a TeV-scale WIMP
can be consistent with the observed DM density,
$\Omega_{\text{DM}}h^2=0.1196\pm0.0031$ (68\% C.L.)~\cite{Ade:2013zuv}.
The WIMP DM is required to be electrically and color neutral; however, 
its electroweak (EW) charges are still scarcely constrained.
The EW charges are characterized by the number of $\SU(2)_L$ components
$n$ and its U(1)$_Y$ hypercharge $Y$. Various DM candidates with
different combinations of $(n,Y)$ have been discussed so far in the
literature \cite{Essig:2007az}.
For instance, an $n=5$ fermion or an $n=7$ scalar multiplet with $Y=0$ 
is often stressed as ``minimal DM'' for its automatic DM stabilization
mechanism \cite{Cirelli:2005uq, *Cirelli:2007xd,*Cirelli:2009uv}. A wino
($n=3$ and $Y=0$) is also a good DM candidate in the supersymmetric
Standard Model (SSM). Moreover, a remnant discrete symmetry resulting
from the grand unified symmetry may give rise to stable DM candidates
charged under the SU(2)$_L\otimes$U(1)$_Y$ gauge interactions
\cite{Kadastik:2009dj, *Kadastik:2009cu, *Frigerio:2009wf}.

In this article, we especially focus on DM with $Y\neq 0$. These
multiplets distinguish themselves from others as they can have
vector-like mass terms. In fact, such possibilities have been already
excluded by the DM direct detection experiments if the DM has only the
gauge interactions; the $Z$-boson exchanging processes
induce the vector-vector coupling between the DM and quarks, which gives
too large DM-nucleus elastic scattering cross sections. However, small
couplings between the Higgs field and the DM, even via
higher-dimensional operators whose cut-off scale $\Lambda$ is much
higher than 
the DM mass, can allow the scenario to evade the constraint.
This is because a non-zero Higgs vacuum expectation value (VEV) makes
the Dirac fermion (complex scalar) split into two Majorana fermions
(real scalars) DM and DM$^\prime$. In this case, the DM cannot interact with
quarks via the vector-vector interaction and thus the DM-nucleon elastic
scattering cross sections are small enough to avoid the
experimental constraints. Therefore, a model with hypercharged DM should
involve some mechanism to generate couplings between the Higgs field and
the DM. One of the most famous examples for such DM is a Higgsino-like
($n=2$ and $Y=1/2$) neutralino in the SSM. Other models are discussed in
Refs.~\cite{Kadastik:2009dj, *Kadastik:2009cu, *Frigerio:2009wf,
Chun:2012yt, *Kang:2013wm}. This kind of model also predicts some
model-dependent physical consequences which are not determined only with
the EW charges of the DM.

The goal of this article is to extract the physical consequences of the
hypercharged DM with coupling to the Higgs boson, in model-independent
manners. Too tiny DM-Higgs coupling reduces the mass splitting between
the DM and DM$^\prime$, $\Delta m$, and the inelastic scattering ${\rm
DM}+ N \to{\rm DM'}+N $ again gives a strong constraint. 
To avoid it, there is a lower bound on the Higgs and DM couplings.
The coupling can also induce the electric dipole moments
(EDMs) of quarks and leptons, as well as the scalar-type DM scattering
with nucleus, which will be probed by present and future progressing
experiments. We mainly focus on the case of fermionic DM, and
show that future experiments have sensitivities to probe this class of DM
candidates and may disfavor the $Y\geq 1$ cases if no signal is
observed. At the end of this article, we briefly comment on the scalar
DM.

\section{Dark matter and HIGGS COUPLINGS}

Let  $\psi_m$ be  the SU(2)$_L$ $n$-tuplet Dirac fermions with
hypercharge $Y>0$. The index $m$ labels the
eigenvalues of $T_3$ with $T_a$ $(a=1,2,3)$ the $n$-dimensional
representation of the generators of the
SU(2)$_L$ gauge group. In the basis, $T_\pm \equiv T_1\pm iT_2$ and
$T_3$ are represented by $(T_\pm)_{lm}=\sqrt{(j\mp m)(j\pm
m+1)}~\delta_{l,m\pm1}$ and $(T_3)_{lm} =m~\delta_{lm}$ with $j\equiv
\frac{n-1}{2}$. We require that the multiplets should contain the
neutral components; the condition reads $Y\leq j$ and $(j-Y)$ being an
integer. Further, the lightest neutral component is assumed to be the
dominant component of DM in the Universe. 
The mass term of the multiplets is given by
\begin{equation}
 \mathcal{L}_{\text{mass}}=-\mu \overline{\psi}\psi ~,
\end{equation}
with $\mu$ taken to be real and positive, without loss of generality. We
assume it to be around the TeV scale in the following discussion. 
Without ultraviolet (UV)-physics effects, the fermions interact with the
Standard Model sector only through the gauge interactions. As discussed
in the Introduction, however, it is required to include the effects to
evade the constraints coming from the DM direct detection
experiments.\footnote{The constraints can be also avoided if $\mu$ is
much higher than the TeV scale. Such a possibility is studied in
Ref.~\cite{Feldstein:2013uha}. } Such
effects are described by the following effective operators that break
the conservation of the fermion number associated with the
multiplets:\footnote{In addition, there is a similar pseudoscalar
operator. However, we find that it plays no role in the following
discussion.  } 
\begin{align}
 \mathcal{L}_{\text{eff}}^{(c)}&= \frac{c_s}{2\Lambda^{(4Y-1)}}
\sum_{M,m,m^\prime}\langle jmjm^\prime |(2Y)M\rangle  [(H)^{4Y}_M ]^*
\overline{\psi^c_m}\psi_{m^\prime}
\nonumber \\&+\text{h.c.}~,
\label{eq:effc}
\end{align}
where $c_s$ is an $\mathcal{O}(1)$ dimension-less constant and $\psi^c$
is the antiparticle field of $\psi$; $\Lambda$ is taken to be real and
positive without loss of generality. $H=(H^+,H^0)^T$ is the Higgs
field; $(H)^k$ is composed of $k$ Higgs fields to form an isospin-$k/2$
object and defined such that its lowest component is given
by $(H^0)^k$; $\langle j mj^\prime m^\prime |JM\rangle $ are the
Clebsch-Gordan coefficients. Notice that from their symmetry properties,
$\langle jmjm^\prime |(2Y)M\rangle 
=(-1)^{2(j-Y)} \langle jm^\prime jm|(2Y)M\rangle$ follows, and thus
the operators in Eq.~\eqref{eq:effc} vanish unless $(j-Y)$ is an
integer. The condition is, however, always satisfied in the present
scenario since we have assumed that the multiplets have the neutral
components. 

In general, the UV-physics effects that induce the above operator also
generate other operators that have lower mass dimensions but give no
contribution to the mass splitting between the neutral components. Among
them, the following dimension-five operators give rise to the dominant
contribution to the low-energy physics:\footnote{Dipole-type operators
are usually suppressed by a loop factor, and thus their contribution is
subdominant.  } 
\begin{align}
 \mathcal{L}_{\text{eff}}^{(d)}&=
\frac{1}{\Lambda}|H|^2 \overline{\psi}(d_s+id_{s5}\gamma_5)\psi
\nonumber \\
&+\frac{1}{\Lambda}(H^*t_a H)\overline{\psi}(d_t+id_{t5}\gamma_5)
T_a\psi
~,
\label{eq:effd}
\end{align}
where $t_a\equiv \sigma_a/2$ with $\sigma_a$ the Pauli matrices, and
the coefficients are dimension-less and of $\mathcal{O}(1)$. In what
follows, we study the phenomenology of these SU(2)$_L$ multiplets in the
presence of the effective operators \eqref{eq:effc} and
\eqref{eq:effd},\footnote{For recent related studies, see
Refs.~\cite{Hisano:2014kua, Nagata:2014wma}. } and discuss the
constraints on $\Lambda$ for each $Y$.

\section{Inelastic Scattering}

As mentioned above, the effective operators in Eq.~\eqref{eq:effc}
generate the mass splitting between the neutral components after the
electroweak symmetry breaking. Once the Higgs field gets a vacuum
expectation value, $\langle
H\rangle =(0,v)^T/\sqrt{2}$ with $v\simeq 246$~GeV, the operators yield
the mass splitting as
\begin{equation}
 \Delta m =\frac{v^{4Y}C_{jY}|c_s|}{2^{(2Y-1)}\Lambda^{(4Y-1)}}~.
\label{eq:massdif}
\end{equation}
Here we define $C_{jY} \equiv \langle jYjY|(2Y)(2Y)\rangle$.

If $\Delta m<\mathcal{O}(100)$~keV, the inelastic scattering of the
DM with a nucleus may occur via the $Z$-boson exchange processes, which is
significantly restricted by the direct detection experiments. The
scattering cross section is  
\begin{equation}
 \sigma_{\text{inel}}=\frac{G_F^2Y^2}{2\pi}[N-(1-4\sin^2\theta_W)Z]^2
M^2_{\text{red}}~.
\end{equation}
Here, $G_F$ is the Fermi constant; $\theta_W$ is the weak mixing angle;
$M_{\text{red}}$ is the reduced mass in the DM-target nucleus system;
$Z$ and $N$ are the numbers of protons and neutrons in the nucleus,
respectively. By using the cross section, we obtain the differential
event rate with the recoil energy $E_R$ in a direct detection
experiment as
\begin{equation}
 \frac{dR}{dE_R}=\frac{N_Tm_T\rho_{\text{DM}}}{2m_{\text{DM}}M^2_{\text{red}}}
 \sigma_{\text{inel}}F^2(E_R)\int_{v_{\text{min}}}^\infty
 \frac{f(v)}{v}dv ~, 
\end{equation}
where $N_T$ is the number of the target nuclei; $m_{\text{DM}}$ and
$m_T$ are the masses of the DM and the nucleus, respectively;
$\rho_{\text{DM}}$ is the local DM density; $f(v)$ is the local DM
velocity distribution; $F^2(E_R)$ denotes a nuclear form factor. The
minimum speed $v_{\text{min}}$ in the integral is given by
\begin{equation}
 v_{\text{min}}=\frac{c}{\sqrt{2m_TE_R}}\biggl(\frac{m_TE_R}{M_{\text{red}}}
+\Delta m\biggr)~.
\end{equation}
Current direct detection experiments have sensitivities to a recoil energy
of $E_R<\mathcal{O}(100)$~keV, and thus the event rate $R$
strongly depends on $\Delta m$ if $\Delta
m<\mathcal{O}(100)$~keV, while the scattering basically never
happens if $\Delta m \gg 1$~MeV. As a consequence, the direct detection
experiments impose a lower limit on the mass difference, which is
interpreted as an upper limit on the scale $\Lambda$ through
the relation in Eq.~\eqref{eq:massdif}. 

\begin{figure}[t]
\begin{center}
 \includegraphics[clip, width = 0.4 \textwidth]{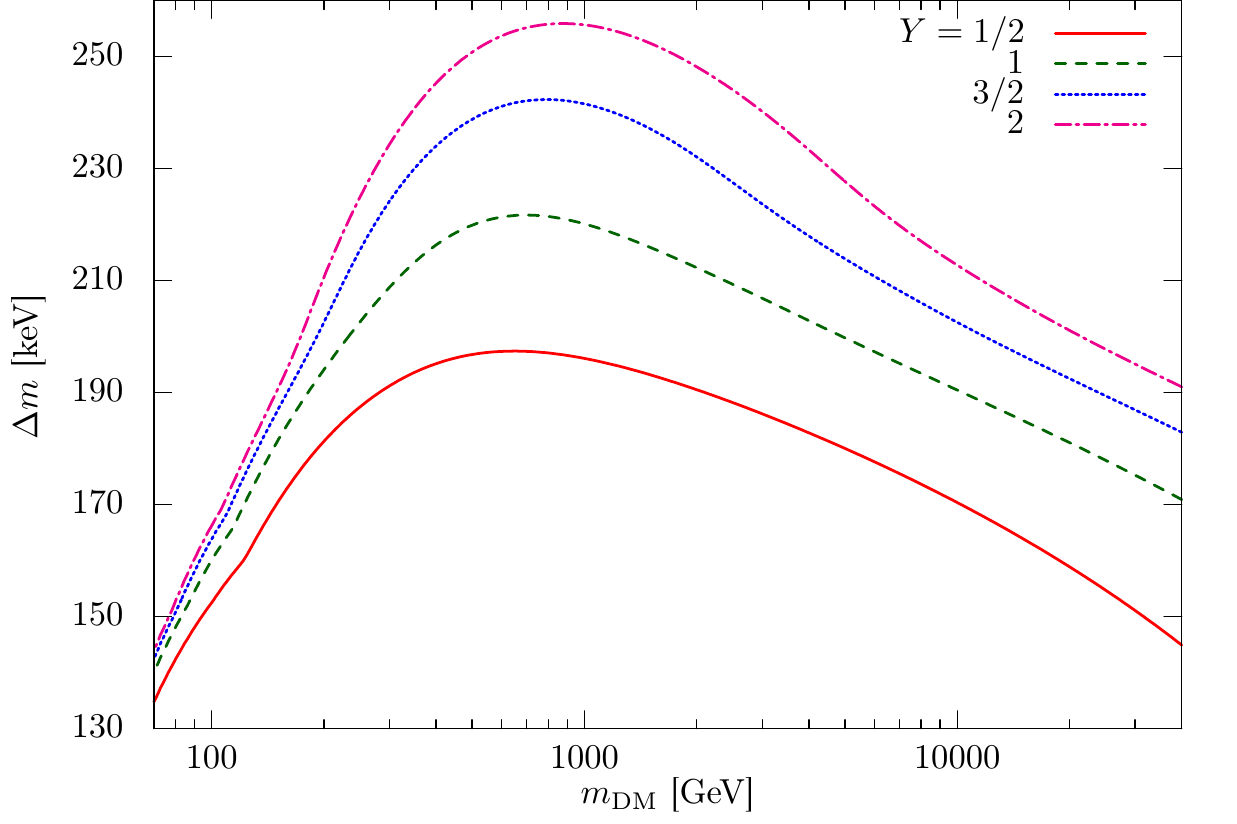}
\caption{Lower bound on the mass splitting $\Delta m$ from the
 inelastic-scattering limits as functions of
 the DM mass $m_{\text{DM}}$. 
 }
\label{fig:inelbound}
\end{center}
\end{figure}

In Fig.~\ref{fig:inelbound}, we show the constraints on the mass
splitting $\Delta m$ coming from the direct detection experiments as
functions of the DM mass $m_{\text{DM}}$. The $Y=1/2,\, 1,\,3/2,\,2$
cases are presented from bottom to top. We combine the results of
XENON10 \cite{Angle:2009xb}, XENON100 \cite{Aprile:2012nq}, and LUX
\cite{Akerib:2013tjd}, and give the lower limits at 90\% C.L. by using a
simple merging and maximum gap method
\cite{Yellin:2002xd,Yellin:2011xf}. We use the same parameters for the
nuclear form factor and the astrophysical DM velocity distribution as
Ref.~\cite{Angle:2009xb}, except for $v_{\rm esc} = 544$~km/s
\cite{Smith:2006ym}.  
Although the constraints strongly depend on these parameters, the limit
$\Delta m>100$~keV is robust in a range of the DM masses shown in the
figure. As a result, we have the upper bounds on $\Lambda$ as
\begin{align}
\Lambda \lesssim (10^9,3\times 10^4, 4\times 10^3)~{\rm GeV}~{\rm for}~
Y=(\frac{1}{2}, 1, \frac{3}{2})~. \label{eq:in_cons}
\end{align}
A larger $Y$ leads to a smaller $\Lambda$, which gives significant
impacts on low-energy observables, as we will see below.

\section{Elastic Scattering}

As we have discussed so far, by considering the
inelastic-scattering processes, we obtain an upper limit on the scale
$\Lambda$ for each multiplet. If the scale is low enough, on the other
hand, the dimension-five operators in Eq.~\eqref{eq:effd} get
significant. The operators induce the DM-nucleon elastic scatterings,
which are again constrained by the direct detection experiments. 
Let us evaluate the cross sections. In the presence of the effective
operators, the DM-quark scalar coupling $f_q$ is induced as
\begin{equation}
 f_q=-\frac{1}{2m_h^2\Lambda}(d_s+\frac{Y}{2}d_t)~,
\label{eq:fq}
\end{equation}
where $m_h$ denotes the mass of the Higgs boson. Here we neglect
the contribution of the operators in Eq.~\eqref{eq:effc}. It is actually
subdominant when $Y\geq 1$. In the following analysis we use
Eq.~\eqref{eq:fq} for the $Y=1/2$ case as well, for brevity. The
inclusion of the contribution is straightforward; see
Ref.~\cite{Nagata:2014wma} for details.  

The DM-quark scalar coupling induces the effective coupling of the DM
with nucleons. The DM-proton coupling is, for instance, given by
\begin{equation}
 \frac{f_p}{m_p}=\sum_{q=u,d,s}f_q f_{T_q}
+\frac{2}{27} \sum_{Q=c,b,t}f_Q f_{T_G}~.
\end{equation} 
Here, $m_p$ is the proton mass, and $f_{T_u}=0.019$,
$f_{T_d}=0.027$, $f_{T_s}=0.009$, and $f_{T_G}\equiv
1-\sum_{q=u,d,s}f_{T_q}$. They are extracted from the recent results of
the lattice QCD simulations \cite{Young:2009zb, *Oksuzian:2012rzb}. In
addition, electroweak gauge boson loop diagrams contribute to the
effective coupling. The contribution is computed as
\cite{Hisano:2011cs,*Hisano:2010fy,*Hisano:2010ct}  
\begin{equation}
 f_p^{\text{EW}} = (n^2-1-4Y^2)f_p^W +Y^2 f_p^Z  ~,
\end{equation}
with $f_p^W\simeq 2.3\times 10^{-11}~\text{GeV}^{-2}$ and $f_p^Z\simeq
-1.1\times 10^{-10}~\text{GeV}^{-2}$. These values scarcely depend on
the DM mass when it is larger than the gauge boson masses. 
The spin-independent (SI) DM-proton elastic scattering cross section
$\sigma^p_{\text{SI}}$ is then given by
\begin{equation}
 \sigma_{\text{SI}}^p=\frac{4}{\pi}M_{\text{red}}^2 f_p^2~.
\end{equation}

\begin{figure}[t]
\begin{center}
 \includegraphics[clip, width = 0.4 \textwidth]{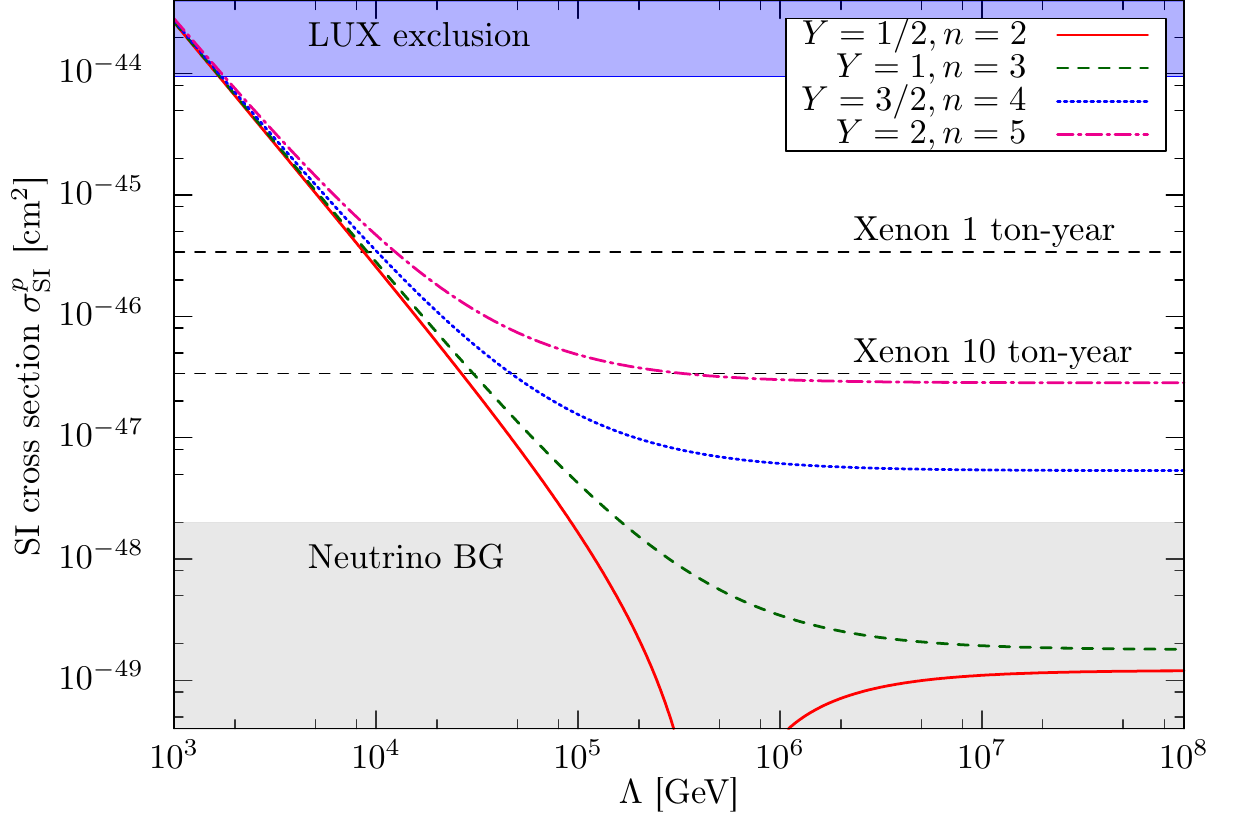}
\caption{DM-proton SI scattering cross sections as functions of
 $\Lambda$. We set $m_{\text{DM}}=1$~TeV, $d_s=1$, and the others
 coefficients to be zero. Blue shaded region represents the present
 bound given by the LUX experiment \cite{Akerib:2013tjd}, while gray
 shaded region indicates the border line below which the neutrino
 background dominates the DM signals \cite{Billard:2013qya}. 
 }
\label{fig:SI}
\end{center}
\end{figure}

In Fig.~\ref{fig:SI}, we show the DM-proton SI scattering cross sections
$\sigma^p_{\text{SI}}$ as functions of $\Lambda$ for some selected
model parameters. The $Y=j=1/2,\, 1,\, 3/2,\, 2$ cases are plotted from
bottom to top. 
We set $m_{\text{DM}}=1$~TeV, $d_s=1$, and the other
coefficients to be zero. The cross sections turn out to be almost
independent of the DM mass. The blue shaded region represents the
current experimental bound given by the LUX Collaboration
\cite{Akerib:2013tjd}. As can be seen, it has already restricted the
region of $\Lambda<$ a few TeV for $m_{\rm DM}=1$~TeV. We also show the
expected sensitivities of the future Xenon-based experiments
\cite{Arisaka:2011eu} in black dashed lines for reference. It is found that
the future experiments can probe $\Lambda = \mathcal{O}(10^{(4-5)})$~GeV,
which is significantly higher than the conditions from the
inelastic-scattering limits (\ref{eq:in_cons}) for $Y\geq 1$. In
addition, larger $Y$ and $n$ tend to yield larger scattering rates via
the electroweak loop contributions, which are independent of
$\Lambda$. Anyway, we see that larger $Y$ scenarios can be relatively
easily tested with future DM detection experiments.

\section{Electric Dipole Moments}

The DM direct detection bounds discussed above are relevant to the
parity-even part of the effective operators. The parity-odd part is, on
the other hand, probed or constrained with the EDMs. The experimental
constraints on the quantities give another lower limit on the scale
$\Lambda$. The EDM of a fermion $f$ is induced at the two-loop
level\footnote{EDMs are, in general, induced also at one-loop level
through the effects of UV physics above the scale $\Lambda$. As long as
$|\mu|\ll \Lambda$ holds, however, such a contribution is subdominant.}
through the so-called Barr-Zee diagrams \cite{Barr:1990vd}, which is
computed as follows: 
\begin{equation}
 d_f=d_f^{h\gamma}+d_f^{hZ}+d_f^{WW}~,
\end{equation}
with 
\begin{align}
 d_f^{h\gamma}&=\frac{e^3Q_fm_fn}{3(4\pi)^4\Lambda\mu}
f_0\biggl(\frac{\mu^2}{m_h^2}\biggr) \nonumber \\[2pt]
&\times[(n^2-1+12Y^2)d_{s5}-Y(n^2-1)d_{t5}]
~, \label{eq:hg} \\[2pt]
d_f^{hZ}&=\frac{eg^2
 m_f n}{12(4\pi)^4\Lambda\mu}(T_f^3-2Q_f\sin^2\theta_W) 
f_1\biggl(\frac{m_Z^2}{m_h^2},\frac{\mu^2}{m_h^2}\biggr) \nonumber \\[2pt]
&\times
[2\{(n^2-1)-12Y^2\tan^2\theta_W\}d_{s5}
\nonumber \\[2pt]
&-Y(n^2-1)(1-\tan^2\theta_W)d_{t5}]
~,\\[2pt]
d_f^{WW}&=\frac{eg^2m_fT_f^3}{6(4\pi)^4\Lambda\mu}Yn(n^2-1)
d_{t5} f_0\biggl(\frac{\mu^2}{m_W^2}\biggr)
~.
\end{align}
Here, $e =|e|$ is the positron charge; $g$ is the SU(2)$_L$ coupling
constant; $m_f$, $Q_f$, and $T_f^3$
are the mass, electric charge in the unit of $e$, and isospin of the
fermion $f$, respectively. The mass functions in the
expressions are 
\begin{align}
f_{0}(r) &= r \int_0^1 dx \frac{1}{r - x(1-x)}\ln\left(
\frac{r}{x(1-x)}
\right), \\[3pt]
f_{1}(r_1, r_2) &= \frac{1}{1-r_1}\biggl[
f_0(r_2)-r_1 f_0\biggl(\frac{r_2}{r_1}\biggr)
\biggr]~.
\end{align}
Currently the electron EDM bound $|d_e|<8.7\times 10^{-29}~e\text{cm}$
by the ACME Collaboration \cite{Baron:2013eja} gives the most stringent
limit on the UV-physics scale. 
For the electron EDM, the $h\gamma$ and $WW$ contributions are dominant.
The prefactor of Eq.~\eqref{eq:hg} is 
\begin{align}
\frac{e^3Q_e m_e}{3(4\pi)^4\Lambda\mu}
f_0
\simeq  -3 \times 10^{-29}~e{\rm cm} \times \left(
\frac{10^{6}~{\rm GeV}^2}{\Lambda \mu}
\right)
\ln\biggl|\frac{\mu}{m_h}\biggr|~.
\end{align}
With $\mathcal{O}(1)$ CP-violating coefficients $d_{s5}$ and $d_{t5}$,
$\Lambda$ less than several TeV is disfavored for $m_{\rm
DM}=\mathcal{O}(1)$ TeV.  

The sensitivity of the EDM measurements is expected to be improved by a
few orders of magnitude in the future \cite{Hudson:2011zz,Vutha:2009ux};
e.g., $|d_e| \sim 10^{-31}~e{\rm cm}$. With the improved measurements
the cut-off scale $\Lambda$ even above the PeV scale can be tested.

\section{Summary and Discussion}

\begin{figure}[t]
\begin{center}
 \includegraphics[clip, width = 0.4 \textwidth]{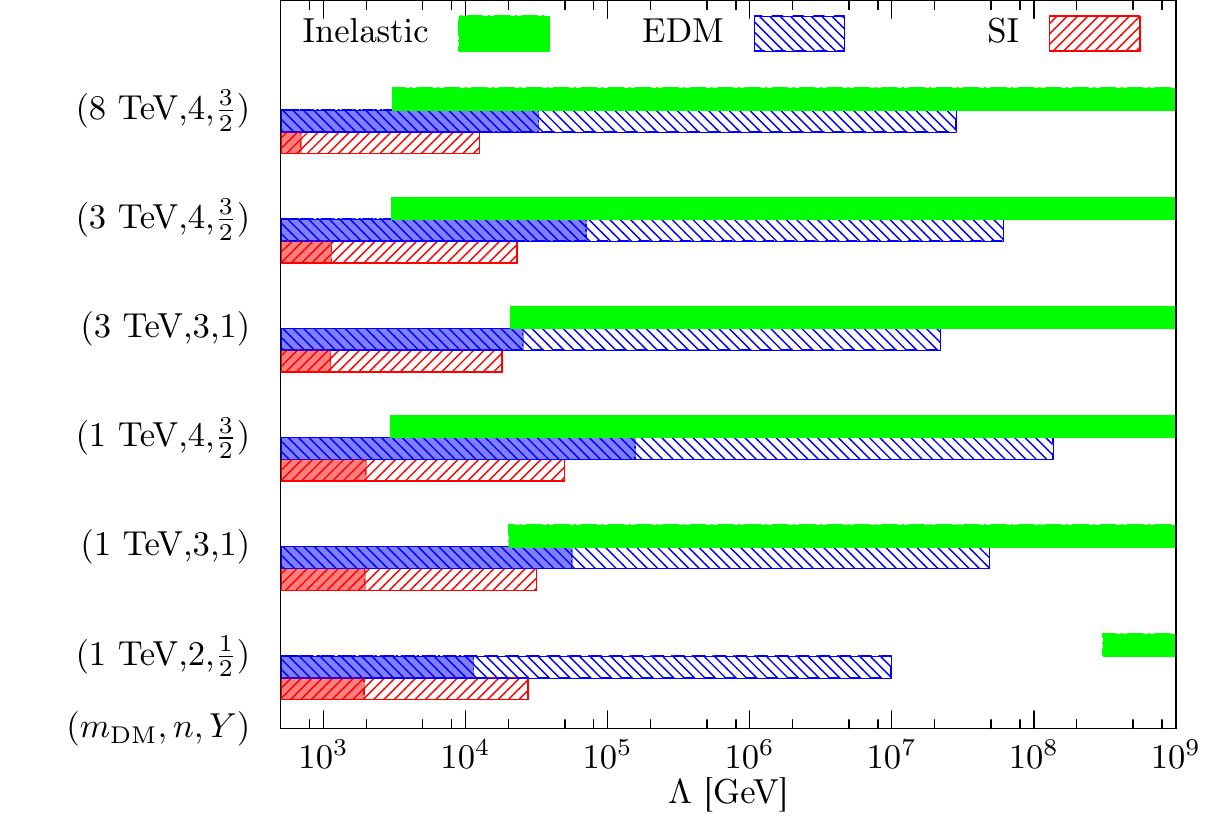}
\caption{Constraints and prospects for the cut-off scale $\Lambda$.
 We set $d_s=d_{s5}=c_s = 1$ and $d_t=d_{t5} =0$.
 Each hatched region filled with (without) the same color shows the
 current constraints (prospects).  
 }
\label{fig:cons_pros}
\end{center}
\end{figure}

We have studied the electroweak interacting DM with non-zero hypercharge.
With the higher-dimensional operators \eqref{eq:effc}, dangerous
$Z$-boson mediated scatterings can be avoided if they give the
mass splitting $\Delta m \gtrsim 100$~keV between the neutral
components. However, other operators with the same cut-off scale
$\Lambda$ may induce large signals for the DM-nucleus elastic
scatterings and/or the EDMs. In Fig.~\ref{fig:cons_pros}, we show the
complementary feature for some selected examples. Here, we take
$d_s=d_{s5}=c_s = 1$ and $d_t=d_{t5} =0$. Each hatched region filled with
(without) the same color shows the current constraints
(prospects). For prospects we refer to the expected reach of a Xenon-based 10
ton-year experiment \cite{Arisaka:2011eu} for the direct detection limits and
$|d_e| = 10^{-31} e$cm for the EDM bounds
\cite{Hudson:2011zz,Vutha:2009ux}. Generally speaking,
a larger $n$ with $Y$ fixed leads to a more severe limit. 
Note that when the cut-off scale $\Lambda$
approaches the DM mass, analyses based on the effective theories
become invalid. Constraints in such a case should be dependent
on each UV model, since it implies an additional sector showing up
around the TeV scale. Generically, however, we may expect
more direct effects on the EDM and DM signals, as well as on the
data in the indirect DM searches and the collider experiments,
coming from this sector. The contributions make the DM more restricted.
Keeping this notice in mind, in Fig.~\ref{fig:cons_pros} we extrapolate
the results computed in effective theories, just for references. 
From this figure, it is found that the DM with $Y\ge 3/2$ are now
strongly disfavored. Even the $Y=1$ cases start to be constrained, and
future experiments can examine the cases. If no signal is observed, only
the $Y=0$ and $1/2$ cases may be the remaining possibilities for the
SU(2)$_L$ charged fermionic DM candidates. 

Finally, we briefly comment on the scalar DM cases. Similarly to the
Dirac fermion DM, a scalar DM with non-zero hypercharge also has the
vector-coupling to $Z$-boson. For the $Y\geq 1$ cases, only
non-renormalizable operators can induce the mass splitting between the
neutral components to avoid the coupling. Thus, in the case of the
$Y\geq 1$ scalar DM, the inelastic bound can give an upper limit on the
UV-physics scale, just like the fermion DM cases. We have $\Lambda
\lesssim (10^5, 4\times 10^3)$~GeV for $Y=1,3/2$, respectively, with
$m_{\text{DM}}=1$~TeV. On the other hand, the
DM-nucleus elastic scattering via the Higgs-boson exchange is induced by
renormalizable interactions, and thus it is not necessarily dependent on
the UV scale. Further, EDMs are not induced and thus play no role in
the scalar DM cases. Nonetheless, when an upper limit on $\Lambda$ is as
low as the DM mass, it indicates the presence of extra particles other
than the DM multiplet around the TeV scale, which provides us various
ways to probe the scalar DM in experiments.

\begin{acknowledgements}
 The work of N.N. is supported by Research Fellowships of the Japan Society
for the Promotion of Science for Young Scientists.
\end{acknowledgements}

\bibliography{ref}

\end{document}